\documentclass[twocolumn,showpacs,amsmath,amssymb,prl]{revtex4}
\usepackage{graphicx}
\usepackage{dcolumn}
\usepackage{bm}
\begin{document}
\title{
Left-handed Ferromagnet
}
\author{Masaru Onoda$^{1,3}$}
\email{m.onoda@aist.go.jp}
\author{Andrei S. Mishchenko$^{1,3}$}
\email{andry.mishenko@aist.go.jp}
\author{Naoto Nagaosa$^{1,2,3}$}
\email{nagaosa@appi.t.u-tokyo.ac.jp}
\affiliation{
$^1$Correlated Electron Research Center (CERC),
National Institute of Advanced Industrial Science and Technology (AIST),
Tsukuba Central 4, Tsukuba 305-8562, Japan\\
$^2$Department of Applied Physics, University of Tokyo, Bunkyo-ku, Tokyo 113-8656, Japan\\
$^3$CREST, Japan Science and Technology Corporation (JST), Saitama, 332-0012, Japan\\
}
\begin{abstract}
The dynamics of the total magnetization in metallic ferromagnet is
studied theoretically taking into account the relativistic spin-orbit interaction.  
It is found that its quantum dynamics is seriously influenced 
by the band crossings near the Fermi energy, 
and sometimes the direction of the precession can be reversed from 
what expected from the commutation relation $[S^{x},S^{y}]= i\hbar S^{z}$ 
($h = 2\pi \hbar$: Planck constant), i.e., the left-handed ferromagnet can be realized.
\end{abstract}
\maketitle

The components of the spin operator satisfy the commutation relation (hereafter we put $\hbar=1$)
\begin{equation}
[S^{\alpha},S^{\beta}]=i\epsilon^{\alpha\beta\gamma}S^{\gamma}
\label{eq:commutator}
\end{equation}
where $\alpha,\beta,\gamma=x,y,z$, and 
$\epsilon^{\alpha\beta\gamma}$ is 
the perfect anti-symmetric tensor with $\epsilon^{xyz} = 1$. 
This means that the spin has the definite handedness, i.e., right-handed nature, and it has been believed that this can never be challenged. 
Therefore, the handedness of the spin is so fundamental and important for the quantum mechanics, 
and it is striking if one can change it.

In solids, the magnetization $\bm{M}$ is the sum of the spin of each 
electrons, i.e., $\bm{M}=\sum_{\bm{r}}\bm{S}_{\bm{r}}$ which 
satisfies also the commutation relation (CR)
$[M^{\alpha}, M^{\beta}] = i 
\epsilon^{\alpha\beta\gamma}M^{\gamma}$.  
In the ferromagnetic state, one can replace $M^{z}$ by the thermodynamic average 
$\langle M^{z} \rangle = M (>0)$ on the r.h.s. of the CR, 
and obtains $[M^{x}, M^{y} ] = i M$ which is analogous to that between 
the position $x$ and the momentum $p$ of a particle. 
Namely, the correspondence $x\to M^{x}$, $p\to M^{y}$ works for the dynamics of the small 
amplitude fluctuation of the transverse magnetization, i.e., spin wave or magnon, in the 
ferromagnetic state. With the easy axis anisotropy energy $D$, the Hamiltonian is given 
by
\begin{equation}
H=D[(M^{x})^2+(M^{y})^2]
\label{eq:D}
\end{equation}
which is equivalent to that of a harmonic oscillator. As can be easily seen from the analogy to the trajectory of $(x,p)$-point in the phase 
space, the direction of the rotation in $(M^{x},M^{y})$-space is
clockwise.
 
The above consideration seems to be quite general and never be 
challenged. However, the influence of the relativistic spin-orbit 
interaction on the CR has not been studied carefully 
to the best of authors' knowledge. 
For the central field of force, the spin-orbit interaction can be written 
as $H_{\mathrm{SO}}=\lambda\bm{S}\cdot\bm{L}$ with $\lambda$ 
being the spin-orbit coupling energy and $\bm{L}$ is the orbital angular 
momentum. Then, the total angular momentum $\bm{J}=\bm{L}+\bm{S}$ is 
the conserved observable, which again satisfies the CR
$[J^{\alpha},J^{\beta}]=i\epsilon^{\alpha\beta\gamma}J^{\gamma}$
\cite{VanVleck}. 
When $\lambda$ is larger than the crystal field as in 
the case of rare-earth ions, the magnetization $\bm{M}$ is given by 
$\bm{M}\sim\sum_{\bm{r}}\bm{J}_{\bm{r}}$ and satisfy the 
same CR.  
Therefore, the above situation does not change. 

In transition metal ions, on the other hand, the orbital angular momentum  
is quenched by the crystal field and band formation, 
and the spin angular momentum mostly contributes to the
magnetization \cite{VanVleck}. 
Furthermore, in the metallic ferromagnet, the 
CR of the magnetization is determined by the quantum mechanical 
Berry phase of the many-body wavefunction and shows sometimes highly nontrivial behavior as shown below.

The repulsion between electrons is the origin of the magnetization, 
because it 
suppresses the double occupancy of electron on each orbital and hence tends to induce 
the spin moment. Typically it can be written as
\begin{equation}
H_{U}=-U\sum_{\bm{r}}\bm{S}_{\bm{r}}\cdot\bm{S}_{\bm{r}},
\end{equation}
where $\bm{S}_{\bm{r}}=\frac{1}{2}\sum_{b}c^{\dagger}_{b\alpha,\bm{r}}
\sigma_{\alpha\beta}c_{b\beta,\bm{r}}$ 
is the spin operator summed over the orbital $b$ on each site $\bm{r}$. 
One can introduce the collective degrees of freedom as 
\begin{equation}
H_{\phi}=U\sum_{\bm{r}}(\bm{\phi}_{\bm{r}}\cdot\bm{\phi}_{\bm{r}}-2\bm{\phi}_{\bm{r}}\cdot\bm{S}_{\bm{r}}).
\end{equation}
Here after integrating over $\bm{\phi}_{\bm{r}}$, we obtain $H_{U}$. This $\bm{\phi}_{\bm{r}}$ 
represents the collective dynamics of magnetization, which is described by the effective action obtained by 
integrating over the electron degrees of freedom. 

Considering the small amplitude vibration of 
$\bm{\phi}_{\bm{r}}$ around the ordered moment, 
the quadratic effective action is enough as given by
\begin{equation}
S[\bm{\phi}]=\sum_{n,\bm{\kappa}}\phi^{\alpha}_{-\bm{\kappa}}
(-i\nu_{n})
\Pi^{\alpha\beta}_{\bm{\kappa}}(i\nu_{n})
\phi^{\beta}_{\bm{\kappa}}(i\nu_{n})
\label{eq:action}
\end{equation}
with $\nu_{n}$ being the Matsubara frequency, and $\bm{\kappa}$ the 
wave vector.
The function $\Pi^{\alpha\beta}_{\bm{\kappa}}(i\nu_{n})$ is given by
\begin{eqnarray}
\Pi^{\alpha\beta}_{\bm{\kappa}}(i\nu_{n})
&=&U\delta^{\alpha\beta}-2U^{2}\chi^{\alpha\beta}_{0\bm{\kappa}}
(i\nu_{n}),
\nonumber\\
\chi^{\alpha\beta}_{0\bm{\kappa}}(i\nu_{n})
&=&\frac{1}{V}\sum_{m,m',\bm{k}}
\frac{f(\epsilon_{m'\bm{k}+\bm{\kappa}})-f(\epsilon_{m\bm{k}})}
{i\nu_{n}+\epsilon_{m\bm{k}}-\epsilon_{m'\bm{k}+\bm{\kappa}}}
\nonumber\\
&&\times
\langle m\bm{k}|S^{\alpha}|m'\bm{k}+\bm{\kappa}\rangle
\langle m'\bm{k}+\bm{\kappa}|S^{\beta}|m\bm{k}\rangle,
\nonumber\\
\end{eqnarray}
where $U$ is the Coulomb repulsion energy, and $|m\bm{k}\rangle$ is the Bloch wavefunction with 
crystal momentum $\bm{k}$ and band index $n$ in the ferromagnetic state. It should be noted 
here that the quantum dynamics of the collective coordinates 
$\bm{\phi}_{\bm{r}}$ is influenced by the 
spin-orbit interaction, which is incorporated in the Bloch wavefunction 
$|m\bm{k}\rangle$. 
Then the next question is how we can see the CR for 
$\bm{\phi}_{\bm{r}}$?  To answer this question, one should remember that the canonical conjugate 
relation $[x,p]=i$ is expressed by the term $p\dot{x}$ in the Lagrangian. Therefore, the linear 
order term in $\nu_{n}$  of 
$\Pi^{\alpha\beta}_{\bm{\kappa}}(i\nu_{n})$ determines the 
CR between components of $\bm{\phi}_{\bm{r}}$. 
As we are interested in the uniform 
magnetization, i.e., $\bm{\kappa}=0$, 
the expression is further simplified for small $\nu_{n}$ as 
\begin{equation}
\chi^{\alpha\beta}_{0\bm{\kappa}=0}(i\nu_{n})
\cong 
\chi^{\alpha\beta}_{0\bm{\kappa}=0}(0)+C^{\alpha\beta}\nu_{n}
\end{equation}
where 
\begin{eqnarray}
C^{\alpha\beta}
&=&\frac{1}{V}\sum_{m,m',\bm{k}}
\frac{f(\epsilon_{m'\bm{k}})-f(\epsilon_{m\bm{k}})}
{(\epsilon_{m\bm{k}}-\epsilon_{m'\bm{k}})^{2}}
\nonumber\\
&&\times
\Im\left(\langle m\bm{k}|S^{\alpha}|m'\bm{k}\rangle
\langle m'\bm{k}|S^{\beta}|m\bm{k}\rangle
\right).
\label{eq:Cxy}
\end{eqnarray}
As discussed above, $C^{xy}$ is especially interesting when 
$\langle \bm{\phi}_{\bm{r}}\rangle \|\bm{e}_{z}$
and small amplitude fluctuation of $(\phi^{x}, \phi^{y})$ is studied.
The denominator
$(\epsilon_{m\bm{k}}-\epsilon_{m'\bm{k}})^{2}$
in Eq.~(\ref{eq:Cxy}) suggests that the (near) degeneracies of the two 
bands contribute dominantly to $C^{xy}$ compared with 
$\chi^{\alpha\beta}_{0\bm{\kappa}=0}(0)$ corresponding to 
the anisotropy energy $D$, which contains only 
$\epsilon_{m\bm{k}}-\epsilon_{m'\bm{k}}$
in the denominator.

Precisely speaking, the handedness of spin dynamics itself
is governed by the signs of 
$\Im[\chi^{-1}_{\bm{\kappa}}(\omega)]^{xy}/\omega$
and $\Re[\chi^{-1}_{\bm{\kappa}}(\omega)]^{ii}$ ($i=x,y$)
at a spin-wave resonance $\omega=\omega_{\mathrm{reso}}$,
where $\chi_{\bm{\kappa}}(\omega)$ is the dynamical magnetic 
susceptibility 
which is approximately given by 
$[\chi^{-1}_{0\bm{\kappa}}(\omega)-2U]^{-1}$. 
However, concerning the lowest excitation at low temperature,
they correspond to those of $C^{xy}$ and $D$, respectively.
It should be also noted that, near the resonance,
the sign of $\Im[\chi^{-1}_{\bm{\kappa}}(\omega)]^{xy}/\omega$
is given by the sign of 
$\Re[\chi^{xy}_{\bm{\kappa}}(\omega)-\chi^{yx}_{\bm{\kappa}}(\omega)]$,
which can be observed by the spin-resolved neutron scattering as discussed later.

Now we demonstrate the nontrivial behavior of $C^{xy}$ using an effective two-dimensional model for 
$t_{2g}$ electrons given as \cite{TAHE},
\begin{eqnarray}
H&=&H_{0}+H_{U}-\sum_{\bm{r}}\mu_{B}\bm{H}\cdot(g_{s}\bm{S}_{\bm{r}}+g_{l}\bm{L}_{\bm{r}}),
\nonumber\\
H_{0}&=&
\sum_{\bm{r}}\biggl[
\lambda_{SO}c^{\dagger}_{\bm{r}}(\bm{l}\cdot\bm{\sigma})c_{\bm{r}}
+\Bigl\{
\sum_{\bm{\mu}=\bm{x},\bm{y}}
(-t_{0})c^{\dagger}_{\bm{r}+\bm{\mu}}c_{\bm{r}}
\nonumber\\
&&
+(\pm t_{1})(c^{\dagger}_{zx,\bm{r}\pm\bm{y}}c_{xy,\bm{r}}
+c^{\dagger}_{yz,\bm{r}\pm\bm{x}}c_{xy,\bm{r}}
)+\mathrm{h.c.}\Bigr\}
\biggr],
\nonumber\\
\label{eq:model}
\end{eqnarray}
where $\mu_B =\frac{|e|}{2m_{e}}$ is the Bohr magneton, $g_{s}=-2$, $g_{l}=-1$, 
and $c^{(\dagger)}_{\bm{r}}$ is the column(row) vector of the 
annihilation(creation) operators of the electron at site $\bm{r}$ with spin and orbital indices, 
and the summations with respect to these indices are abbreviated when they are not 
explicitly indicated. The spin and orbital angular momenta are represented by 
$\bm{S}_{\bm{r}}=\frac{1}{2}c^{\dagger}_{\bm{r}}\bm{\sigma}c_{\bm{r}}$ and  
$\bm{L}_{\bm{r}}=c^{\dagger}_{\bm{r}}\bm{l}c_{\bm{r}}$, respectively. 
The matrix $\bm{l}$ is originally $l = 2$
representation of SU(2), and is projected to $t_{2g}$-space in the above model. The transfer 
integrals are determined by the oxygen p-orbitals between the two transition metal ions; 
$t_{0}$ is nonzero even for the perfect perovskite structure, while $t_{1}$ becomes nonzero due to 
the shift of the oxygen atoms out of plane in the $z$-direction. 

In Ref.~\cite{TAHE}, the anomalous Hall effect (AHE) of the model 
Eq.~(\ref{eq:model}) has been studied. 
There, the sign of the transverse conductivity $\sigma^{xy}$ represents 
the direction of the rotational motion of electrons. 
An essential observation is that $\sigma^{xy}$ represents the 
topological nature, 
Berry phase, of the Bloch wavefunction, and is mostly determined by 
the band crossings acting 
as magnetic monopoles in momentum space. 
The expression of $C^{xy}$ in Eq.~(\ref{eq:Cxy}) is related to
that of $\sigma^{xy}$ under the substitution 
$\bm{S}\leftrightarrow\bm{J}_{e}$, where 
$\bm{J}_{e}$ the electric current \cite{TAHE}. 
Therefore, we also expect the 
sign change of $C^{xy}$ for magnon, which is actually the case as shown below.

\begin{figure}[hbt]
\includegraphics[scale=0.35]{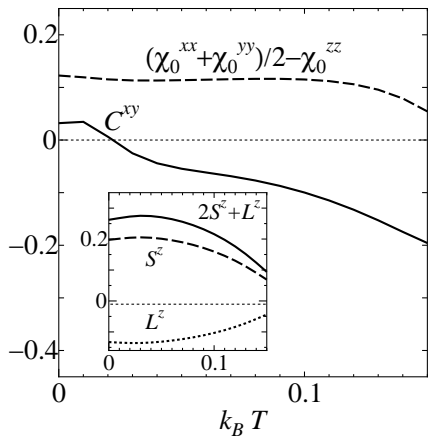}
\caption{
$C^{xy}$ and magnetic anisotropy as functions of temperature.
It is noted that the magnetic anisotropy,
which is measured by 
$[\chi^{xx}_{0\bm{\kappa}=0}(0)+\chi^{yy}_{0\bm{\kappa}=0}(0)]/2-\chi^{zz}_{0\bm{\kappa}=0}(0)$ 
is almost constant in the ferromagnetic phase,
and the sign of $C^{xy}$ changes at low temperature.
This means the interchange between the left- and right-handedness.
Inset shows the temperature dependence of the ordered spin $(S^z)$ and
orbital $(L^z)$ moments.
} 
\label{fig:Cxy}
\end{figure}

Starting from the ferromagnetic ordered state 
$\langle\bm{\phi}_{\bm{r}}\rangle=\bm{\phi}_{\mathrm{mf}}$, 
we obtain the action Eq.~(\ref{eq:action}) for 
$\delta\bm{\phi}_{\bm{r}} =\bm{\phi}_{\bm{r}}-\bm{\phi}_{\mathrm{mf}}$
 by integrating over the electrons. The calculated $C^{xy}$  as a 
function of temperature $T$ is shown in Fig.~\ref{fig:Cxy} ($t_{1} = 0.5$, $\lambda_{SO} = 0.4$, $U = 2.26$, 
chemical potential $\mu =-1.88$ in the unit of transfer integral $t_{0} = 1$),
where we have solved the mean field equation at each $T$. It is seen in Fig.~\ref{fig:Cxy} that the sign of 
$C^{xy}$ changes from negative (right-handed) to positive (left-handed) as the temperature 
is lowered. This means the \textit{effective} sign reversal of the commutator $[M^{x}, M^{y}]$.
This is analogous to the physics of negative-refraction metamaterials,
where negative refractive index appears effectively and causes novel phenomena,
while the microscopic refractive index is not negative
nor does not violate causality \cite{metamaterials}.
The spin anisotropy energy $D$ in Eq.~(\ref{eq:D}) which is given by  
$[\chi^{xx}_{0\bm{\kappa}=0}(0)+\chi^{yy}_{0\bm{\kappa}=0}(0)]/2-\chi^{zz}_{0\bm{\kappa}=0}(0)$  
apart from a constant factor, on the other hand, 
remains almost constant and positive \cite{tiny-H}. 
 
The quantity $C^{xy}$ is closely related to the topological nature of Bloch functions similarly 
to the  transverse conductivity 
$\sigma^{xy}$ \cite{TKNN}.  Actually, at $T=0$, it is written as                              
$C^{xy}=\Im\langle\frac{\partial\Psi}{\partial h^{x}}|\frac{\partial\Psi}{\partial h^{y}}\rangle$
where $\Psi$ is the many-body ground state wavefunction of electrons 
under the magnetic field 
$\bm{h}=(h^{x},h^{y},h^{z})$ applied to the spins \cite{Niu}. 
The Berry curvature is known to be singularly 
enhanced near the band degeneracy acting as "a magnetic monopole".
Here we introduce an effective Hamiltonian 
which describes the nearly degenerate bands of up and down spins. 
$H=\epsilon_{\bm{k}}+\bm{b}_{\bm{k}}\cdot\bm{\sigma}$,
where $\epsilon_{\bm{k}}$ and $\bm{b}_{\bm{k}}$ are arbitrary functions of $\bm{k}$
and magnetization.
The contribution to $C^{xy}$ from these bands is estimated by
\begin{eqnarray}
C^{xy}_{\mathrm{Dirac}}&=&\frac{1}{V}\sum_{\bm{k}}
\left[f(\epsilon_{\bm{k}-})-f(\epsilon_{\bm{k}+})\right]
\frac{b^{z}_{\bm{k}}}{(2|\bm{b}_{\bm{k}}|)^3},
\end{eqnarray}
where $\epsilon_{\bm{k}\pm}=\epsilon_{\bm{k}}\pm|\bm{b}_{\bm{k}}|$.
In the case where the Fermi level is close to this nearly degenerate point,
the above contribution becomes large as the separation between the bands, 
$2|\bm{b}_{\bm{k}}|$, decreases.
When the dominantly minority spin state is 
lower in energy than that of the majority spin state, 
i.e., $b^z_{\bm{k}} >0$,
this enhanced contribution has the opposite sign compared to 
that expected from the original CR Eq.~(\ref{eq:commutator}). 
This occurs only when both the minority and majority bands have finite density of states 
at the Fermi energy, and does not occur for the half-metallic ferromagnets.

Now we turn to experimental consequences of the sign change in $C^{xy}$. One of the 
most direct detection of left-handedness is by means of polarization analysis of 
magnetic neutron scattering. Energy and angle resolved polarization of neutrons $\bm{P}'$ 
scattered by magnetic interaction is proportional to the space and time Fourier transform 
of the statistical average \cite{Marshall} 
\begin{eqnarray}
\bm{P}'&\propto&
g^{2}\Bigl[
\langle
\bm{S}^{(\perp)}_{\bm{r}}(0)\{\bm{P}\cdot\bm{S}^{(\perp)}_{\bm{r}'}(t)\}
+\{\bm{P}\cdot\bm{S}^{(\perp)}_{\bm{r}}(0)\}\bm{S}^{(\perp)}_{\bm{r}'}(t)
\rangle
\nonumber\\
&&-\bm{P}
\langle
\bm{S}^{(\perp)}_{\bm{r}}(0)\cdot\bm{S}^{(\perp)}_{\bm{r}'}(t)
\rangle
-i\langle
\bm{S}^{(\perp)}_{\bm{r}}(0)\times\bm{S}^{(\perp)}_{\bm{r}'}(t)
\rangle
\Bigr].
\nonumber\\
\label{eq:neutron}
\end{eqnarray}
Here $\bm{P}$ is polarization of the incident beam, 
$\bm{S}^{(\perp)}_{\bm{r}}(t) =\tilde{\bm{\kappa}}\times\{\bm{S}_{\bm{r}}(t)\times\tilde{\bm{\kappa}}\}$, 
$\bm{r}$ and $\bm{r}'$ are the 
spin coordinates, $g$ is the gyromagnetic ratio or the Lande splitting factor, 
and $\tilde{\bm{\kappa}}$ is a unit vector along the momentum transfer $\bm{\kappa}$. 
The last term in Eq.~(\ref{eq:neutron}), proportional to 
$\langle\bm{S}^{(\perp)}_{\bm{r}}(0)\times\bm{S}^{(\perp)}_{\bm{r}'}(t)\rangle$, 
has two characteristic features distinguishing it from the all other ones. 
First, it does not depend on the polarization of the incident beam $\bm{P}$. 
Second, this average distinguishes between left- and right-handed 
coordinates
since the direction of the resultant vector $\bm{c}$ of vector product $\bm{a}\times\bm{b}$
is related to the vectors $\bm{a}$ and $\bm{b}$ by the right-hand rule. 
Therefore, to test the violation of the clockwise direction of the 
rotation of spin in the ferromagnet, one has to study polarization of scattered neutron 
beam when the incident beam is unpolarized. In this case the resulting polarization is 
determined purely by the average 
$\langle\bm{S}^{(\perp)}_{\bm{r}}(0)\times\bm{S}^{(\perp)}_{\bm{r}'}(t)\rangle$.

For usual ferromagnet the expression Eq.(\ref{eq:neutron}) can be 
considerably simplified. 
For unpolarized incident beam the resultant polarization of the 
scattered beam is 
proportional to $\bm{P}'\propto 
2\tilde{\bm{\kappa}}g^{2}(\tilde{\bm{\kappa}}\cdot\tilde{\bm{\eta}})
\langle S^{x}(0)S^{y}(t)-S^{y}(0)S^{x}(t)\rangle$.
Here $\tilde{\bm{\eta}}$ is a unit vector along $z$-direction. 
Note, the sign of the polarization of the scattered beam depends just 
of the sign of the average $\langle S^{x}(0)S^{y}(t)-S^{y}(0)S^{x}(t)\rangle$. 
For the particular case of the Heisenberg antiferromagnet the polarization of the scattered neutrons 
with unpolarized scattered beam is 
$\bm{P}_{\pm}' \propto \pm 2\tilde{\bm{\kappa}}(\tilde{\bm{\kappa}}\cdot\tilde{\bm{\eta}})
/[1+(\tilde{\bm{\kappa}}\cdot\tilde{\bm{\eta}})^{2}]$,
where upper sign refers to the process of 
magnon creation and the lower one to its annihilation. 
Hence, if the average $\langle S^{x}(0)S^{y}(t)-S^{y}(0)S^{x}(t)\rangle$
changes its sign due to the spin-orbit coupling, one can observe "wrong" signs of polarizations 
when magnons are created and annihilated.  
Note, the strongest polarization observed for the momentum transfer parallel or 
anti-parallel to the magnetization axis.  Therefore, in terms of the spin-resolved neutron 
scattering, we can obtain the information about the following value 
$
A_{\bm{\kappa}}(\omega)
=
\int^{\infty}_{-\infty}dt e^{i\omega t}
\Im\langle S^{x}_{-\bm{\kappa}}(0)S^{y}_{\bm{\kappa}}(t)\rangle
\propto
\frac{\Re[\chi^{xy}_{\bm{\kappa}}(\omega)-\chi^{yx}_{\bm{\kappa}}(\omega)]}
{1-e^{-\frac{\omega}{k_{B}T}}}
$.
To detect the direction of the precession, we need to know the sign 
of $\Re[\chi^{xy}_{\bm{\kappa}}(\omega)-\chi^{yx}_{\bm{\kappa}}(\omega)]$
at the lowest resonance. 
In addition, there is another important aspect of the neutron scattering spectrum associated 
with the sign change of $C^{xy}$. Namely, as the $C^{xy}$ approaches zero, there splits off the 
bound state from the continuum in $A_{\bm{\kappa}=0}(\omega)$, which has the left-handed direction of 
precession. This state becomes lower in energy, and eventually becomes 
the lowest-energy peak as the sign of $C^{xy}$ changes. Therefore, this multiple peak structure is an 
important signature observed near the transition between right- and left-handedness of 
the magnon modes as expected in the neutron scattering experiment.

\begin{figure}[hbt]
\includegraphics[scale=0.35]{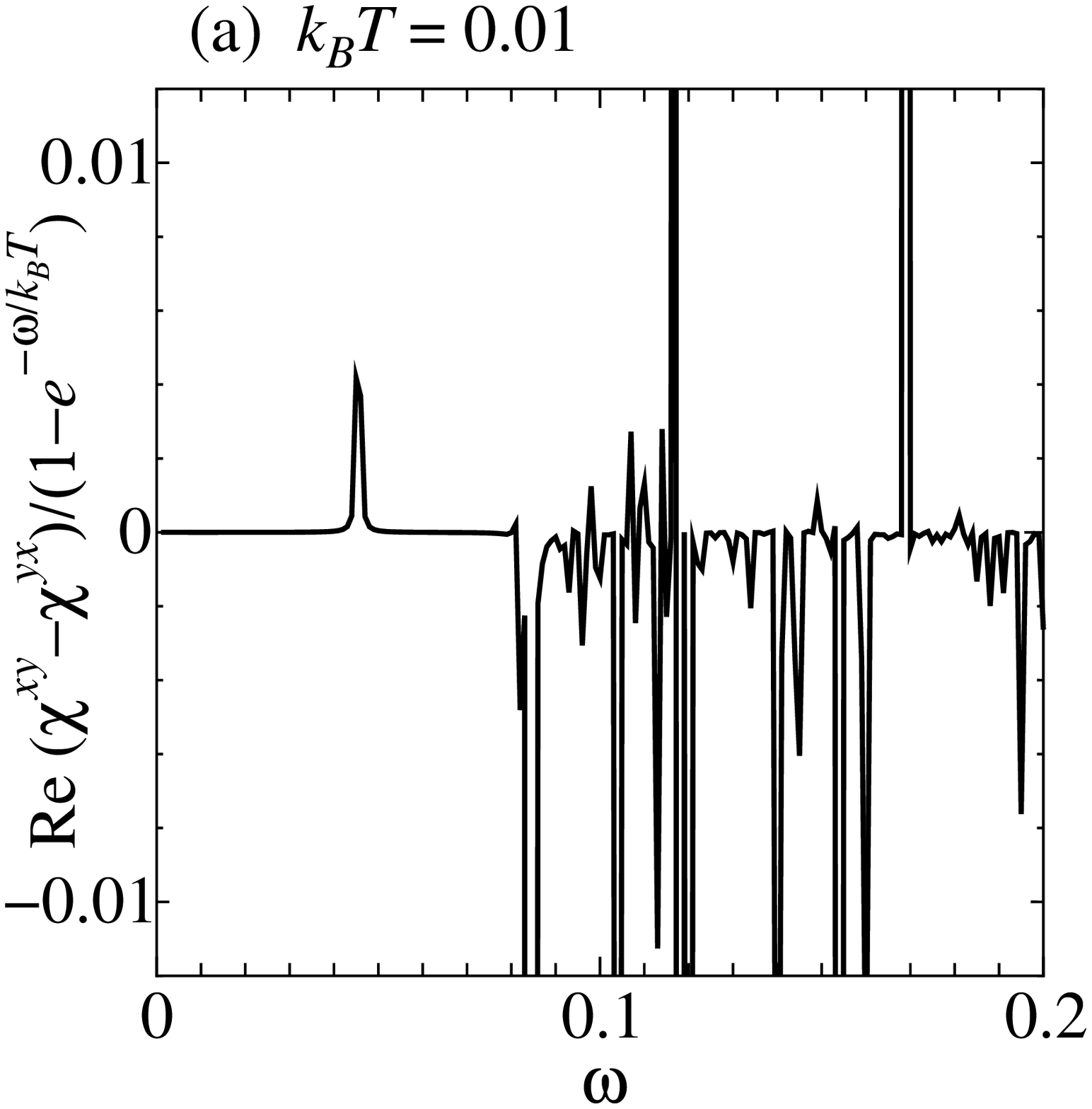}
\includegraphics[scale=0.35]{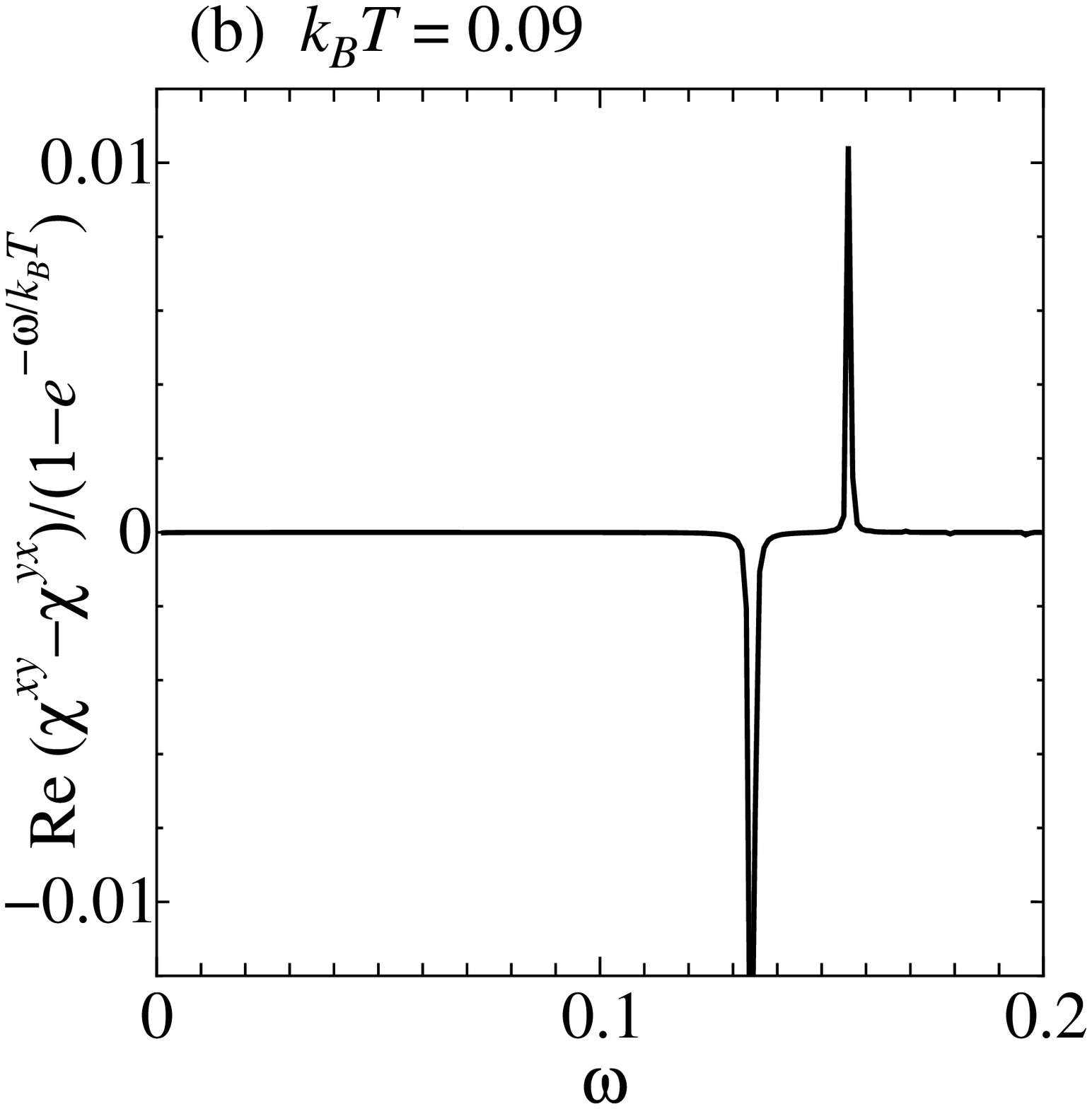}
\caption{
Temperature dependence of the off-diagonal correlation 
obtained by spin-polarized neutron scattering.
The interchange of two spin-wave spectra occurs around
$k_{B}T \sim 0.08$.
The sign of the lowest-energy resonance represents
the left($+$) and right($-$) handedness.
} 
\label{fig:neutron}
\end{figure}

Figure~\ref{fig:neutron} shows the explicit calculation for the model in Eq.~(\ref{eq:model}).  
We can clearly see the interchange of the lowest and the next excitations with $\bm{\kappa}=0$ 
around $k_{B}T \sim 0.08$. 
The lowest-energy excitation corresponds to the left-handed spin motion below $k_{B}T \sim 0.08$, 
e.g. at $k_{B}T \sim 0.01$ in Fig.~\ref{fig:neutron}(a), while the lowest one is right-handed at 
$k_{B}T \sim 0.09$ in Fig.~\ref{fig:neutron}(b).

There are several other experimental techniques which are sensitive to the 
direction of the spin precession in the effective magnetic field, i.e.,
precession $d\bm{S}/dt=g_{\mathrm{eff}}\mu_{B}\bm{H}\times\bm{S}$ 
which direction in external or internal magnetic field of a 
ferromagnet is determined by effective gyromagnetic ratio $g_{\mathrm{eff}}$. 
The most traditional one is the electronic spin resonance (ESR) technique 
with rotating magnetic field \cite{ESR} and 
the most recent advanced one is the magneto-optical polarimetric analysis where the real 
space trajectory of the magnetization dynamics can be detected \cite{Ogasawara,Vomir,Stanciu}. 
However, both techniques are sensitive to the sign of effective $g$-factor. 
Note, the change of the sign of CR, signaling on the left-handedness, 
is an exotic situation. 
To the contrast, negative value of the $g$-factor of effective spin 
Hamiltonian is a frequent 
situation when an ion is influenced by a crystal field from the 
neighboring ions of the lattice \cite{Sabinsky}. 
Hence, we conclude that the polarization analysis of the scattered 
neutrons is a rather unique possibility to trace the left-handedness since the 
polarization of the 
scattered beam Eq.(\ref{eq:neutron}) is determined just by the sign 
of the correlation function and does 
not depend on the sign of the effective $g$-factor. 

In conclusion, we have theoretically shown that the left-handed magnet can be 
realized and does not contradict with the commutation relation for the spin operators. 
The conditions for this novel possibility is that (i) the orbital angular momentum is 
basically quenched but still the spin-orbit interaction affects the spin dynamics, (ii) both 
majority and minority spins have the finite density of states, and (iii) the band 
degeneracy occurs near the Fermi energy and gives the dominant contribution to the 
Berry phase. These conditions are satisfied in the metallic ferromagnets including the 
transition metal ions, and most preferably in two dimensional systems. For example, 
candidate substances are listed up as follows; the transition metal intermetallic compounds Co$_{2}$Pd(Pt)$_{4}$ \cite{Guo}, 
CoPt$_{3}$ \cite{Iwashita}, and the series of ferromagnetic ruthenates                   
Sr$_{1-x}$Ca$_{x}$RuO$_{3}$ \cite{Mathieu}. 
Even though the left-handedness is not realized, we expect the 
nontrivial temperature dependence of the magnon dynamics there. 
Detailed first-principles calculations and experimental search for 
this left-handed magnet is highly desired and left for future 
investigations.

The authors thank Y. Endoh, G.Y. Guo, and Y. Tokura for fruitful discussions.
The work was supported by Grant-in-Aids under the Grant numbers 15104006, 
16076205, and 17105002, and NAREGI Nanoscience Project from 
the Ministry of Education, Culture, Sports, Science, and Technology,
RFBR 07-02-00067a.


\begin{thebibliography}{99}
\bibitem{tiny-H}
Here we applied a $z$-directional magnetic field
$\mu_{B}H = -10^{-3}$ in the unit of $t_{0} = 1$
just to stabilize the out-of-plane magnetic order up to high temperatures. 
This is not essential for the left-handed nature, and is only 
for the demonstration of the two-magnon branches at higher temperature
as discussed for Fig.~\ref{fig:neutron}.
\end{thebibliography}
\end{document}